\documentclass[aps,preprint, superscriptaddress,nofootinbib]{revtex4}
\usepackage{amsmath}
\usepackage{graphicx}
\usepackage{subfigure}
\usepackage{amssymb}
\usepackage{xcolor}
\usepackage{multirow}
\usepackage{cancel}
\usepackage{color}
\usepackage{epsfig}
\usepackage{ulem}
\usepackage{listings}
\usepackage{xcolor}
\usepackage{float}
\usepackage{slashed}

\usepackage{tikz}
\usepackage{pgffor}							% For repeating patterns

\usepackage[colorlinks,citecolor=blue]{hyperref}
\usepackage{amsmath}%\ge=\geqslant; \le=\leqslant
\usepackage{wrapfig}

\newcommand{\be}{\begin{equation}}
\newcommand{\ee}{\end{equation}}
\newcommand{\beq}{\begin{equation}}
\newcommand{\eeq}{\end{equation}}
\newcommand{\bea}{\begin{eqnarray}}
\newcommand{\eea}{\end{eqnarray}}
\newcommand{\besp}{\begin{equation}\begin{split}}
\newcommand{\eesp}{\end{split}\end{equation}}

\newcommand{\Dfbd}{\mathord{\buildrel{\lower3pt\hbox{$\scriptscriptstyle\leftrightarrow$}}\over {D}_{\mu}}}

\def\0{\textbf{0}}
\def\1{\textbf{1}}
\def\2{\textbf{2}}
\def\3{\textbf{3}}
\def\4{\textbf{4}}
\def\5{\textbf{5}}
\def\6{\textbf{6}}
\def\7{\textbf{7}}
\def\8{\textbf{8}}
\def\9{\textbf{9}}

\begin{document}
\title{Direct detection of finite-size dark matter via electron recoil }

\author{Wenyu Wang}
\email{wywang@bjut.edu.cn}

\author{Wu-Long Xu}
\email{wlxu@emails.bjut.edu.cn}
\affiliation{Faculty of Science, Beijing University of Technology, Beijing  100124, P. R. China}

\author{Jin Min Yang}
\email{jmyang@itp.ac.cn}
\affiliation{CAS Key Laboratory of Theoretical Physics, Institute of Theoretical Physics, Chinese Academy of Sciences, Beijing 100190, P. R. China}
\affiliation{ School of Physical Sciences, University of Chinese Academy of Sciences,  Beijing 100049, P. R. China}

\begin{abstract}	
In direct dark matter (DM) detection via scattering off the electrons, the momentum transfer plays a crucial role. Previous work showed that for self-interacting DM, if the DM particle has a size (the so-called puffy DM), the radius effect could dominate the momentum transfer and become another source of velocity dependence for self-scattering cross section. In this work we investigate the direct detection of puffy DM particles with different radii through electron recoil. We find that comparing with the available experimental exclusion limits dominated by the mediator effect for XENON10,  XENON100  and XENON1T, the constraints on the puffy DM-electron scattering cross-section become much weaker for large radius DM particles. For small-radius DM particles, the constraints remain similar to the point-like DM case. 
\end{abstract}
\maketitle

\tableofcontents
%\newpage

\section{Introduction}\label{sec1}
More than eighty percent of the matter in our universe is considered to be dark matter (DM), which does not participate in electromagnetic interaction~\cite{Planck:2018vyg,Springel:2006vs,Bahcall:1999xn}.  Via gravitational interaction,  the existence of DM has been strongly supported by evidences which include Galaxy and Cluster rotation curves, gravitational lensing, the bullet cluster and the Cosmic Microwave Background radiation  \cite{Begeman:1991iy,Dyson:1920cwa,Clowe:2006eq}.  Meanwhile, various new physics models have been proposed to explain DM.   One  of the popular DM candidates is the Weakly Interacting Massive Particle (WIMP) which provides the so-called WIMP miracle due to its mass around the electroweak scale and its thermal freeze-out naturally satisfying the observed relic density \cite{Lee:1977ua}.    However, the signature of WIMP has not observed yet  in the laboratory \cite{Jungman:1995df,Bertone:2018krk}. Further properties of the DM particle, such as  its spin, mass and size need to be studied on the theoretical and experimental sides in the future \cite{Planck:2018vyg}.

In  traditional  direct detection experiments of DM,  the scale of the DM mass is focused above the $\rm GeV$ scale \cite{GAMBIT:2017zdo} and so far no evidences have been found.  Below the $\rm GeV$ scale, the sensitivity of direct detection technique via observing the signals of nuclear recoils from the elastic
DM-nucleus scattering drops quickly \cite{PandaX-II:2018xpz,XENON:2018voc,LUX:2017ree}.  As a result, more attention and technological development is being paid on searching for signatures of scattering between DM and the electrons. For example, in the two-phase xenon time projection chamber (TPC), the xenon atom is ionized and the electron is recoiled. At the same time, other atoms may also be ionized by the recoiling electrons. These recoiling electrons are then drawn through the xenon gas region via an external electric field, which results in the emission of scintillation (S2) signals \cite{Essig:2015cda,Essig:2012yx,Essig:2011nj,Essig:2017kqs,Chen:2015pha,Bloch:2020uzh,Derenzo:2016fse,Blanco:2019lrf,Pandey:2018esq,Liu:2021avx,XENONCollaborationSS:2021sgk,Chavarria:2022par}. However, from a theoretical perspective, it is indistinguishable whether the ionized signals are caused by DM scattering off the electrons or by DM scattering off the nucleus through the Migdal effect \cite{Ibe:2017yqa,Essig:2019xkx,Wang:2021oha,Knapen:2020aky,Liang:2020ryg}.

From a model-building perspective, the idea that the DM particle has a size provides a solution to the small scale structure anomalies in the universe, which include the core-cusp problem and the Too-Big-to-Fail problem, {\it etc} ~\cite{Chu:2018faw}.  This scenario is called the puffy self-interacting DM (SIDM) model, in which the radius effect  of DM particles are considered as another source of velocity dependence of the 
self-scattering  cross section $\sigma/m_{\rm DM}$ \cite{Wang:2023xii,Wang:2021tjf,Chu:2019awd,Laha:2015yoa}.  In fact,  in many new physics models DM is assumed to be the bound state of point-like particles which may be some dark meson, nucleus, atom, {\it etc}.  Meanwhile, such finite-size DM can be also considered as one kind of SIDM paradigm \cite{Laha:2013gva,Cline:2021itd,Tsai:2020vpi}. Given that the puffy DM could lead to a wide range of phenomena and the radius effect could dominate the scattering momentum transfer \cite{Hardy:2015boa, Acevedo:2021kly}, we need to examine the direct detection of the finite-size DM via electron recoil, which is the aim of this work. 

This work is organized as follows. In Sec. II,  the finite-size DM-electron ionization rates  are studied.  In Sec. III,  the numerical results and analyses are described.  Section IV  gives our conclusions.

\section{Puffy DM-electron ionization rates }\label{sec2}
In the scattering process of two point-like particles, the momentum transfer plays a crucial role in calculating the scattering cross-section.  When a particle has a finite size, the scattering process becomes considerably more complex. The reason is that if the momentum transfer is much smaller than the inverse radius of the particle, it can be treated as a point-like particle, and the internal structure of a puffy particle cannot be probed. However, when the momentum transfer is larger than the inverse radius of the DM particle, i.e., $1<qR_{\rm DM}(\leq10^{-3} m_{\rm DM}R_{\rm DM})$, the internal structure can be possibly measured, and a form factor needs to be introduced ~\cite{Engel:1991wq,Feldstein:2009tr}.  Especially, when the mediator mass is very heavy, the conventional cross section will be dominated by the size effect for puffy DM scattering. In our study, the shape of puffy DM particle is taken as the dipole form.
The form factor is introduced as $F(\vec{\bf q})=\int d\vec{\bf r}e^{i\vec{\bf q}\cdot \vec{\bf r}}\rho(\vec{\bf r})$ with the puffy DM  charge density profiles $\rho\textcolor{red}{(r)}=\frac{e^{-r/r_{\rm DM}}}{8\pi r_{\rm DM}^3}$. Thus,
the form factor describing the comprehensive momentum transfer caused by the spatial distribution of puffy DM charge  in the bulk is expressed as 
\be\label{form}
F_{\rm RDM}(q)=\frac{1}{(1+r^2_{\rm DM}q^2)^2},
\ee 
where $r_{\rm DM}$ is the radius of the DM particle and $q$ is the momentum transfer in the scattering. 
When the electron is scattered by puffy DM, the matrix element is corrected by  multiplying this form factor relative to the point particle case \cite{Wang:2023wrx,Wang:2021tjf}.

Scattering between an electron and a DM particle with finite size can be treated in the same way as for the point-like DM case, both mathematically and physically.  The target electrons are considered as single-particle states of an isolated atom, which are calculated by the numerical Roothaan-Hartree-Fock (RHF) bound wave functions \cite{Bunge:1993jsz}. The velocity-averaged differential cross section of the DM scattering off electron is given as \cite{Chu:2018faw,Wang:2023xii,Essig:2017kqs,Essig:2012yx}
\be\label{cross}
\frac{d\left<\sigma_{ion}^{nl}\right>}{d\ln E_{R}}=\frac{\bar{\sigma}_{e}}{8\mu_{\chi e}^{2}}\int qdq|f_{ion}^{nl}(k',q)|^{2}|F_{\rm DM}(q)|^{2}|F_{\rm RDM}(q)|^{2}\eta(v_{\rm min}),
\ee
where $\bar{\sigma}_e$ and $\mu_{\chi e}=m_{\chi}m_e/(m_{\chi}+m_e)$ are  respectively the reduced cross section at fixed momentum transfer $q=\alpha m_e$ and the reduced mass among the DM and electron, $E_R$ and $q$ denote respectively the electron recoil energy and momentum transfer, 
$F_{\rm DM}(q)$ is  the $q$ dependence of the matrix element  called as  the form factor of the DM,  
$f_{ion}^{nl}(k',q)$ is the form factor for ionization electron in the $(n,l)$ shell and  can be expressed as 
\be\label{ion}
|f_{ion}^{nl}(k',q)|^{2}=\frac{k'^{3}}{4\pi^{3}}\times2\underset{n,l,l'm'}{\sum}|\langle f|e^{i{\bf q}_{e}\cdot{\bf x}_{i}}|i\rangle|^{2},
\ee 
with the sum being over all electron shells and the momentum $k'=\sqrt{2m_eE_R}$ \cite{Pandey:2018esq,Catena:2019gfa,Griffin:2021znd}. The inverse mean speed $\eta(v_{\rm min})$, which obeys the  Maxwell-Boltzmann velocity distribution of DM with the circular velocity $v_0=220~ \rm km/s$  and the escaped velocity $v_{\rm esc}=544~ \rm km/s$, is the  function of the minimum velocity $v_{\rm min}$ of the DM particle required for the scattering.  In this scattering process, the DM particle has an incoming velocity of $v$, and the outgoing momenta of the DM particle and the electron are ${\bf p'}_{\chi}$ and $m_ev_e$, respectively. According to momentum conservation, we have 
\be\label{mom}
{\bf q}=m_{\chi}{\bf v}-{\bf p'}_{\chi} =m_e{\bf v_e}.
\ee
The energy transferred into the electron can be obtained via energy conservation 
\be\label{ene}
\Delta E_e =\frac{1}{2}m_{\chi}v^2-\frac{|m_{\chi}v-q|^2}{2m_{\chi}}
={\bf v \cdot q}-\frac{{\bf q^2}}{2m_{\chi}}.
\ee 
Meanwhile, $\Delta E_e=E_b+E_R$ is related to the binding energy $E_b$.  Thus, the minimum velocity  is 
\be\label{minv}
v_{\rm min}=\frac{|E_b^{nl}|+E_R}{q}+\frac{q}{2m_{\chi}} .
\ee 
Then the complete expression of  the inverse mean speed function $\eta(v_{\rm min})$ (The final piecewise function form is as shown in appendix B of the Ref.~ \cite{Essig:2015cda}) is written as 
\be\label{eta}
\eta(v_{\rm min})=\int_{v_{\rm min}}\frac{d^3v}{v}f_{\rm MB}(\vec{v})\Theta(v-v_{\rm min}),
\ee 
where $f_{\rm MB}(\vec{v})$ (see also the Ref. \cite{Maity:2020wic,Radick:2020qip,Savage:2008er}) is  the standard Maxwell-Boltzmann velocity distribution in the galactic rest frame with the   hard cutoff  velocity $v_{\rm esc}=600 ~\rm km/s$ \cite{Dehnen:1997cq,Smith:2006ym}, typical  velocity of the halo DM $v_0=230~ \rm km/s$, and $v_E=240~\rm km/s$ for the Earth velocity relative to the DM halo, adding(subtracting) $15~\rm km/s$ for the Earth velocity in June (December).
In the Earth’s frame,  the velocity distribution can be written as 
\be 
f_{\rm MB}(\vec{v})=\frac{1}{Nv_0^3\pi}e^{-\frac{|\vec{v}+\vec{v}_E|^2}{v_0^2}}\Theta(v_{\rm esc}-|\vec{v}+\vec{v}_E|), 
\ee 
where the normalization factor is 
\be 
N=\sqrt{\pi}{\rm Erf}\left(\frac{v_{\rm esc}}{v_0}\right)-2\left(\frac{v_{\rm esc}}{v_0}\right)e^{-\left(\frac{v_{\rm esc}^2}{v_0^2}\right)}
\ee 
The differential ionization rate is obtained as 
\be \label{diff}
\frac{dR_{ion}}{d\ln E_{R}}=N_{T}\frac{\rho_{\chi}}{m_{\chi}}\sum_{nl}\frac{d\left<\sigma_{ion}^{nl}v\right>}{d\ln E_{R}},
\ee 
where $N_T$ is the number of target atoms and the local matter density $\rho_{\chi}=0.4~\rm GeV/cm^3$.
Additionally,  from a field-theoretical perspective,  a predictive benchmark model should be implemented.  We assume that the Dirac fermion DM particle $\chi$ interacts with the electron via exchanging a dark photon mediator $A'$ with mass $m_v$. The Lagrangian of this simplified mode is 
\be\label{lag}
\mathcal{L} \supset  g_{\chi}\bar{\chi}\gamma_{\mu}\chi A^{'\mu}+g_e\bar{e}\gamma_{\mu}eA^{'\mu},
\ee 
where $g_{\chi}$ and $g_e$ are the coupling constants. 

The scattering cross section between puffy DM and electron can be written as a product of the reduced cross section $\bar{\sigma}_e$,  the momentum-dependent factor $F_{\rm DM}(q)$ of the  point particle case  and  $F_{\rm R DM}(q)$  in which the radius effect will be included:
\be\label{sigm}
\sigma_e=\bar{\sigma}_e|F_{\rm DM}|^2 |F_{\rm RDM}|^2,
\ee 
where $\bar{\sigma}_e$ is taken as 
\be\label{red}
\bar{\sigma}_e=\mu^2_{\chi e}\frac{1}{16\pi m^2_{\chi}m^2_{e}} \overline{|M_{\rm point}(q=q_0)|^2}, 
\ee 
with the reference momentum $q_0=\alpha m_e$ and $\alpha=g_{\chi}^2/4\pi$. 

Finally, the form factor $F_{\rm DM}(q)$, in which the momentum dependence is from the mediator effect, can be written as \cite{Feldstein:2009tr}
\beq\label{formf}
\begin{split}
|F_{\rm DM}(q)|^2&=\frac{\overline{|M(q)_{\rm point}|^2}}{|M_{\rm point}(q=q_0)|^2}\\
&=\frac{(m_{v}^{2}+q_{0}^{2})^{2}(8m_{e}^{2}m_{\chi}^{2}-2q^{2}(m_{e}+m_{\chi})^{2}+q^{4})}{(m_{v}^{2}+q^{2})^{2}(8m_{e}^{2}m_{\chi}^{2}-2q_{0}^{2}(m_{e}+m_{\chi})^{2}+q_{0}^{4})}.
\end{split}
\eeq 
From  Eqs. (\ref{form},\ref{sigm},\ref{formf}) we see that when the radius of DM particle is very small, the momentum dependence is dominated by the mediator, namely $F_{\rm DM}(q)=1$ for a heavy dark photon or $F_{\rm DM}(q)=\alpha^2m_e^2/q^2$ for a ultra-light dark photon, which is in agreement with the study in ~\cite{Essig:2017kqs}. It can be expressed as 
\begin{align}\label{smallr}
F_{\rm DM} (q)F_{\rm RDM} (q)\quad
\underrightarrow{r\rightarrow0}\quad F_{\rm DM} (q)& \approx \begin{cases}
1 &{\rm for ~a ~massive ~mediator}\\
\hspace{2cm}\  & \ \\[-4.1mm]
\alpha^2m_e^2/q^2& {\rm for ~a ~massless ~mediator} \\
\end{cases}
\end{align}
If the DM particle has a sizable radius, $F_{\rm RDM} (q)\sim 1/(r^4_{\rm DM}q^4)$, the total form factor can be rewritten approximately as
\begin{align}\label{bigr}
F_{\rm DM} (q)F_{\rm RDM} (q)\quad
\underrightarrow{r\rightarrow \infty}\quad & \begin{cases}
\frac{1}{r^4_{\rm DM}q^4} &{\rm for ~a ~massive ~mediator}\\
\hspace{2cm}\  & \ \\[-4.1mm]
\frac{\alpha^2m_e^2}{r^4_{\rm DM}q^6} & {\rm for ~a ~massless ~mediator} \\
\end{cases}
\end{align}
This implies a suppression of total momentum transfer for a large radius of DM. In the following,  the effects from such a finite radius will be studied numerically.

\section{Numerical analysis and results}\label{sec3}
\begin{figure}[!htbp]
	\centering
	\includegraphics[width=7.5cm]{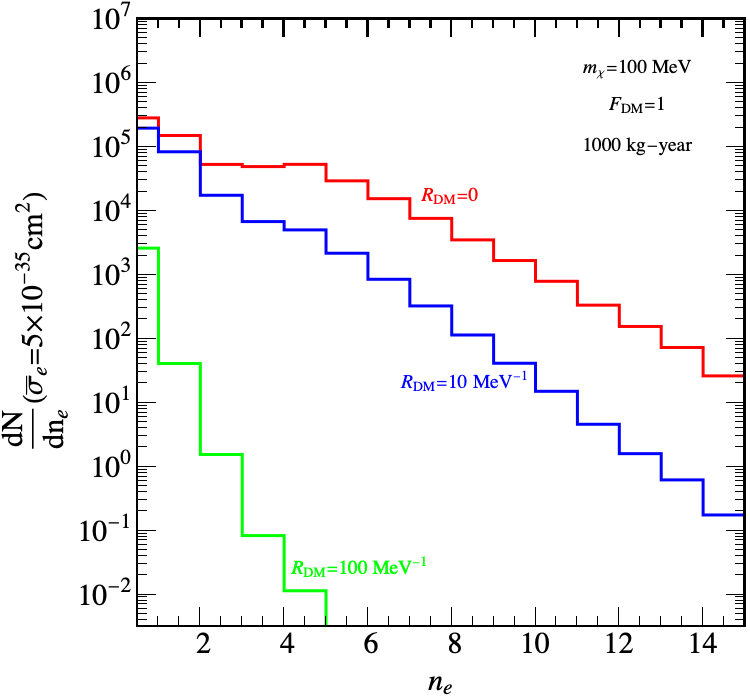}
	\includegraphics[width=7.5cm]{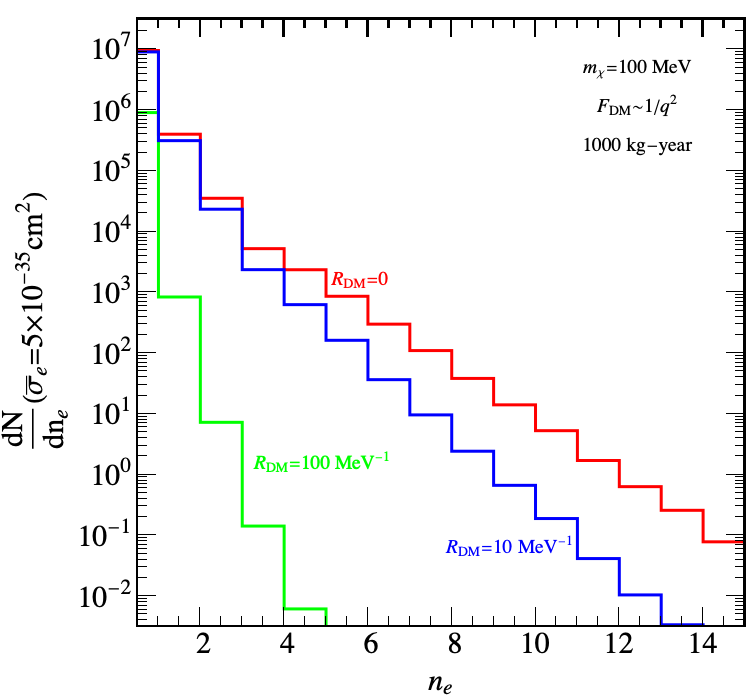}
 \vspace{-.6cm}
	\caption{Spectrum of expected number of events of   different radius DM-electron scattering in $F_{\rm DM}=1 $ case with $\bar{\sigma}_e=5\times 10^{-35}~ \rm{cm^{2}}$ (left panel) and  $F_{\rm DM}=q_0^2/q^2$ case with $\bar{\sigma}_e=5\times 10^{-39}~\rm{cm^{2}}$ (right panel) in xenon.  The mass of DM is chosen as $m_{\chi}=100~ \rm MeV$ and the exposure is $\rm 1000 kg-year$. }
	\label{fig1}
\end{figure}

As in ~\cite{Essig:2012yx}, the constraints on the scattering cross section between puffy DM and electron can be obtained by using the experimental data. Firstly, the observed electron number $n_e$ and the scintillation photon number $n_{\gamma}$ can be deduced from the electron recoil energy $E_R$. The step energy 
$W= 13.8 ~\rm eV$ (the average energy required to create a single quantum \cite{Shutt:2006ed}) can produce the  primary quantum number $n^{(1)}$ defined as $n^{(1)}= \rm Floor(E_R/W)$ where Floor is the rounded down function. The probability of the initial electron recombining with an ion is considered as $f_R=0$ and the fraction for the observed  electrons is $f_e=0.83$. The corresponding uncertainties are set as $0<f_R<0.2$, $12.4~\rm eV <W<16~\rm eV$ and $0.62<f_e<0.91$.  The shells exciting the photos are $\rm (5s, 4d, 4p, 4s)$ with energy $(13.3, 63.2, 87.9, 201.4)  \rm ~eV$ respectively. The additional quantum numbers under the photnionization effect are $n^{(2)}=(0,4,6-10,3-15)$ shown in Table \ref{table1} \cite{Essig:2017kqs}. And the total quantum number $n^{(1)}+n^{(2)}$ obey the binomial distribution.   In FIG. \ref{fig1},  the recoil spectra  of electron  are shown for $F_{\rm DM}=1 $ and  $F_{\rm DM}=q_0^2/q^2$ with different radius of DM.  Comparing with the point particle case $R_{\rm DM}=0$, in which the momentum transfer is dominated by the mediator effect,  
we see that for a small radius $R_{\rm DM}=\rm 10~ MeV^{-1}$ the suppression from the radius effect of puffy DM is not sizable in the range of electron number $1-15$ for $F_{\rm DM}=1 $ and $F_{\rm DM}=q_0^2/q^2$. However, for a large radius $R_{\rm DM}=\rm 100 ~MeV^{-1}$ the suppression from the radius effect is significant and the momentum transfer is dominated by the radius effect.  Beyond that, when the number of ionization electrons is larger than 3, the recoil spectrum is very small.  
%%%%table 1
\begin{table}[!ht]
	\centering
\begin{tabular}
{|c|c|c|c|c|c|}	\hline
shell&$5p^6$&$5s^2$&$4d^{10}$&$4p^6$&$4s^2$ \\ \hline
~~Binding Energy [eV]~~&~~12.6~~&~~25.7~~&~~75.5~~&~~163.5~~&~~213.8~~ \\ \hline
Additional Quanta&0&0&4&6-10&3-15 \\ \hline
\end{tabular}	
	\caption{The binding energy and  additional quanta for the shells of Xenon $(5p^6, 5s^2, 4d^{10} 4p^6, 4s^2)$.  }\label{table1}
\end{table}

\begin{figure}[!htbp]
	\centering
	\includegraphics[width=5.2cm]{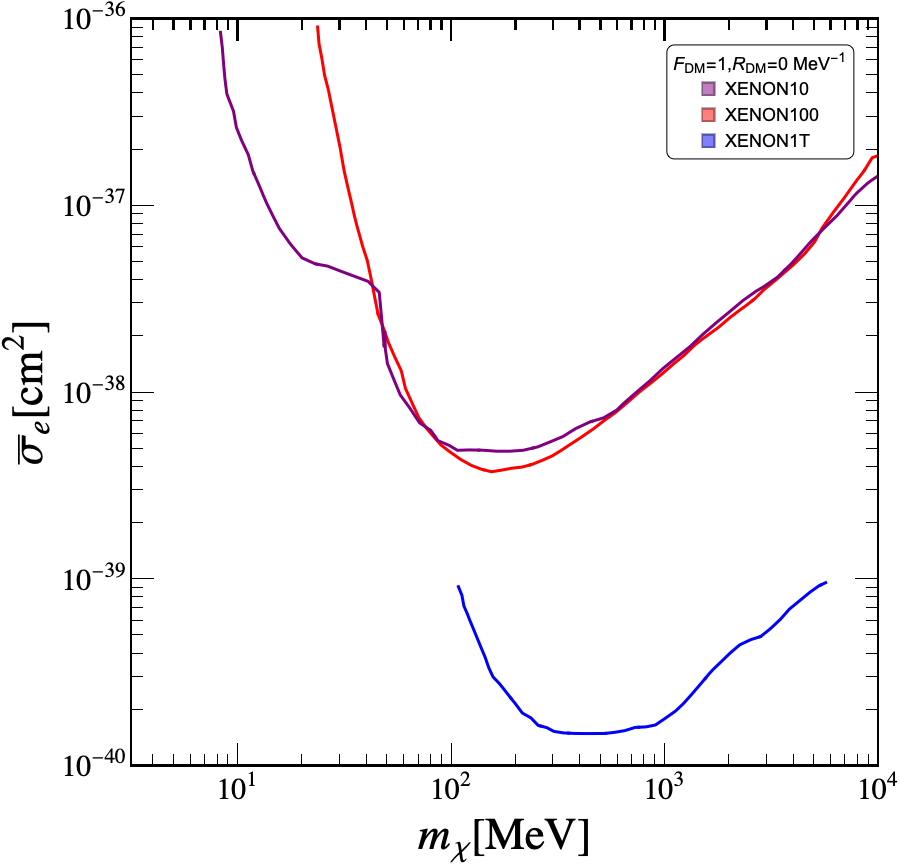}
	\includegraphics[width=5.2cm]{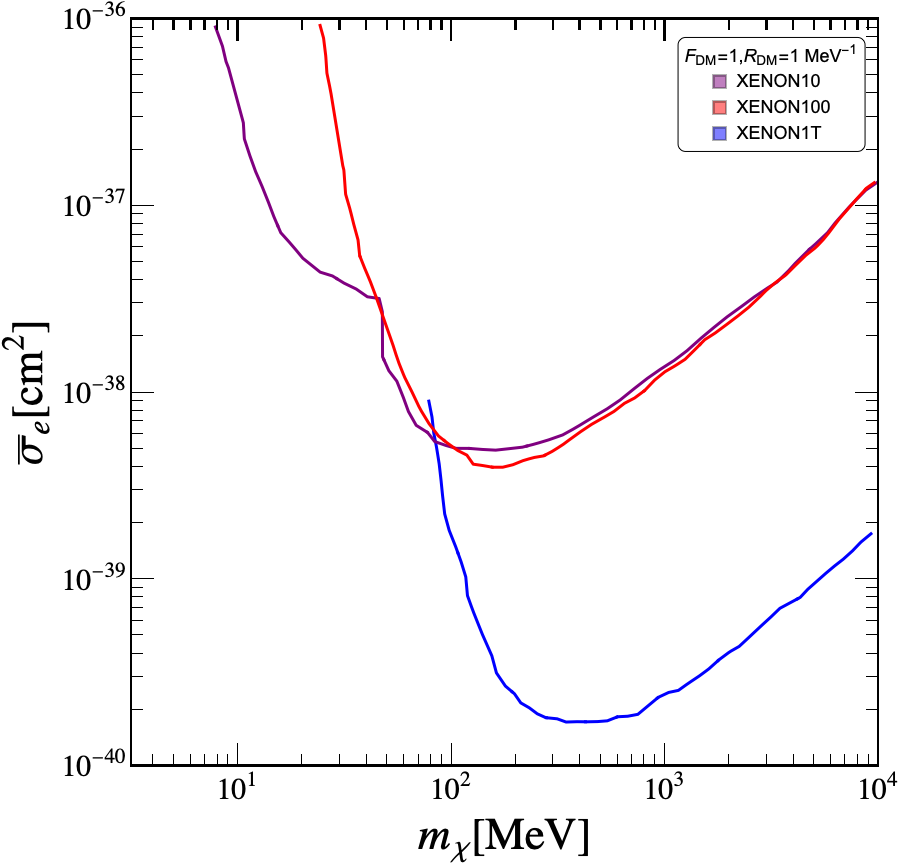}
		\includegraphics[width=5.2cm]{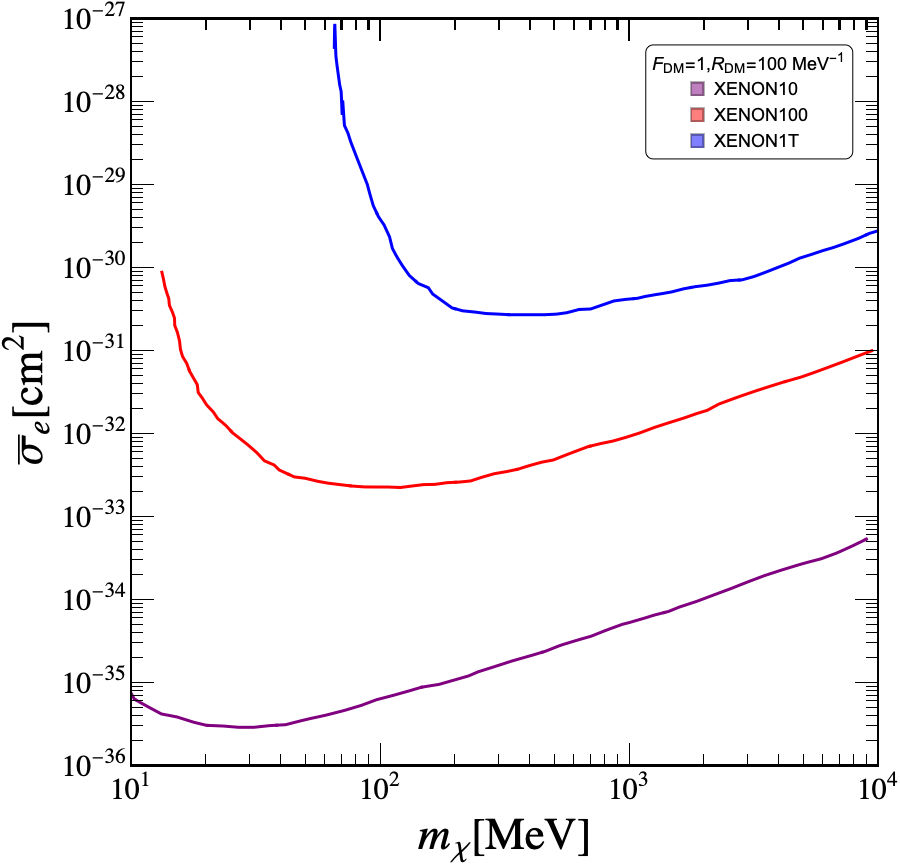}\\
			\includegraphics[width=5.2cm]{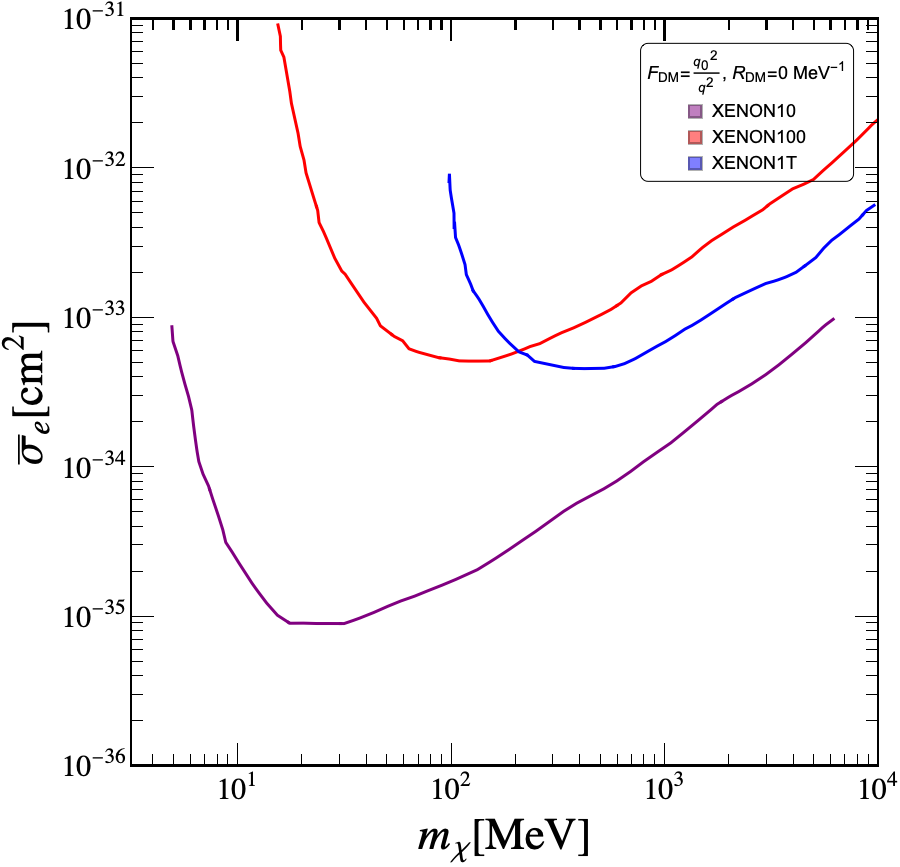}
			\includegraphics[width=5.2cm]{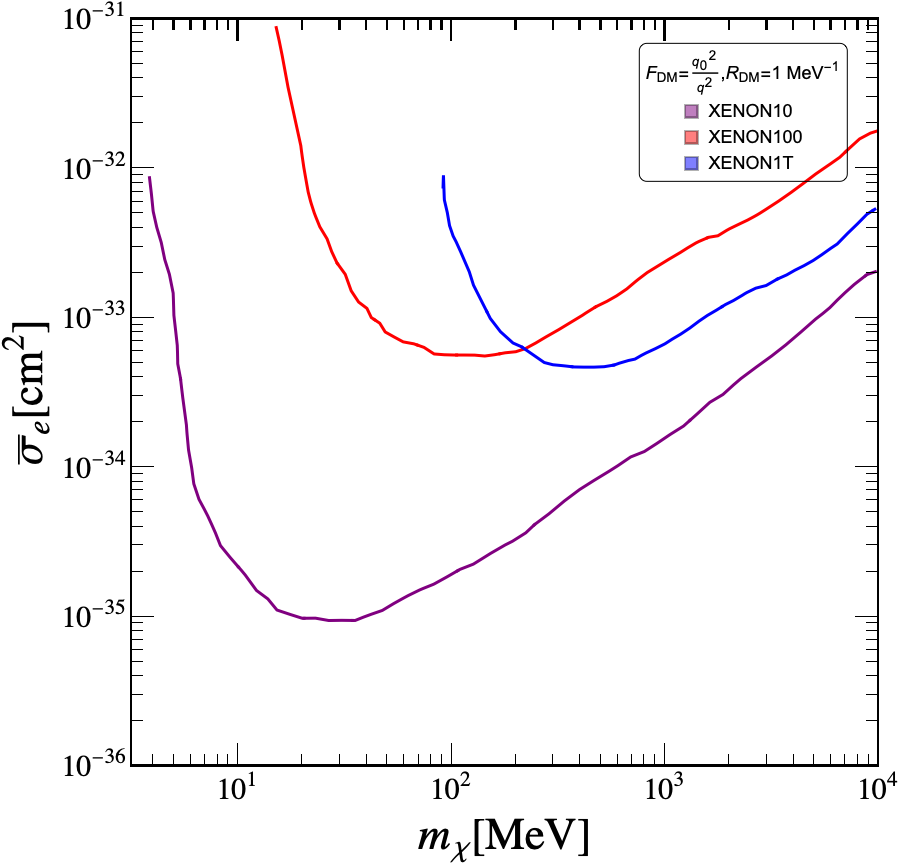}
			\includegraphics[width=5.2cm]{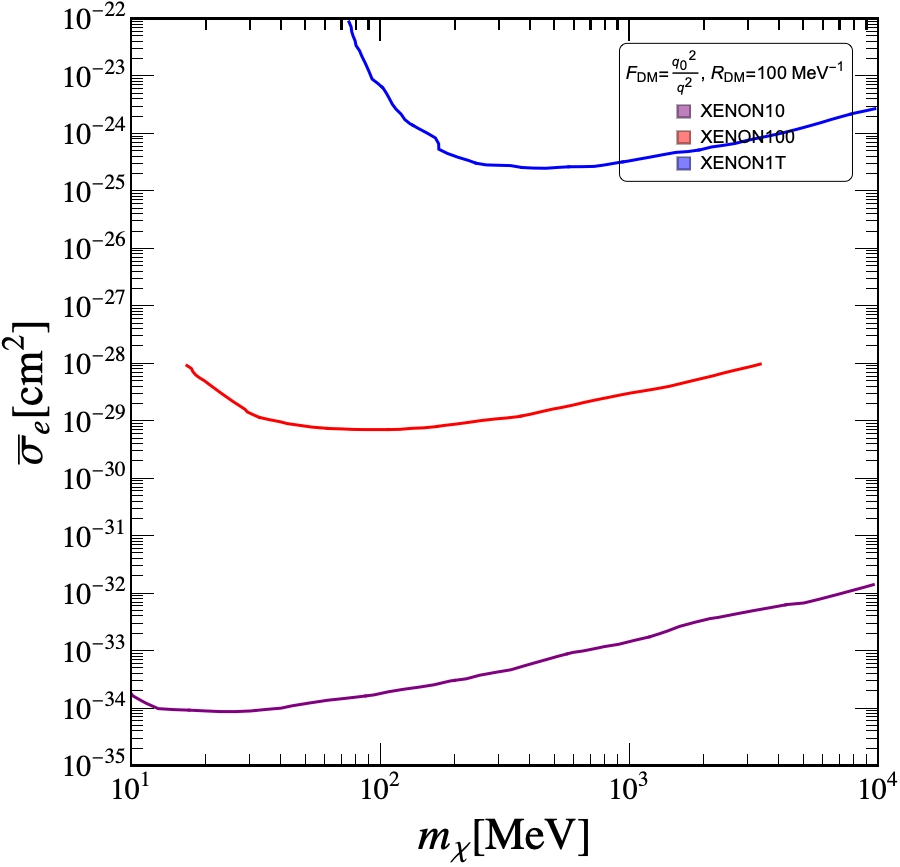}\\
		
	\caption{The limits on the different radius DM-electron scattering cross section for $F_{\rm DM}=1 $ (upper panels) and  $F_{\rm DM}=q_0^2/q^2$ (lower panels) in the case of $R_{\rm DM}= 0$, $R_{\rm DM}= 1~ \rm MeV^{-1}$ and $R_{\rm DM}=100~ \rm MeV^{-1}$ from XENON10 data (purple),  XENON100 data (red) and XENON1T data (blue). }
	\label{fig2}	
\end{figure}

Then in a given event the signal of $n_e$ will be converted into photo-electrons (PEs) which are produced by photomultiplier tuber  to observe the experiment results.  For an event with $n_e$, the PE number can be obtained  via a Gaussian function with mean $n_e\mu$ and width $\sqrt{n_e\sigma}$. Here $\mu=27, 19.7, 11.4$ and $\sigma=6.7, 6.2, 2.8$ for XENON10, XENON100, XENON1T, respectively. We set the PE bins as in \cite{Essig:2012yx, XENON100:2011cza, XENON:2019gfn}. Note that the PE bin for the XENON1T in our work is set as the range of  $[165, 275]$.  Finally, to compare with the theoretical value, the experimental data of XENON10 (15-kg-days) \cite{Essig:2012yx}, XENON100 (30-kg-years) \cite{XENON:2016jmt,XENON100:2011cza} and XENON1T (1.5 tones-years) \cite{XENON:2019gfn} are applied.  FIG. \ref{fig2} shows that the limits   of  the different radius DM-electron scattering cross section for $F_{\rm DM}=1 $ (upper three panels) and  $F_{\rm DM}=q_0^2/q^2$ (lower three panels) in the case of $R_{\rm DM}= 0$, $R_{\rm DM}= 1~ \rm MeV^{-1}$ and $R_{\rm DM}=100~ \rm MeV^{-1}$ respectively from XENON10 data (purple),  XENON100 data (red) and XENON1T data (blue).
  Comparing  the limits with $R_{\rm DM}= 0$  and $R_{\rm DM}= 1~ \rm MeV^{-1}$,  we can infer that for a very small radius of DM particle, the momentum transfer is much smaller than the inverse radius and the puffy DM can be seen as a point-like particle. In this case the momentum transfer is dominated by the mediator effect, namely $F_{\rm DM}(q)F_{\rm RDM}(q) \sim F_{\rm DM}(q)$, and the excluded limits are similar to the case of point DM particle.  Instead,  for large radius puffy DM, the momentum transfer is dominated by the radius effect,  namely $F_{\rm DM}(q)F_{\rm RDM}(q) \sim F_{\rm RDM}(q)$, and the excluded limits are much weaker than for the point DM particle.  Thus,  when the DM particle has a large size, the momentum transfer from  the radius effect  cannot be ignored  and it weakens the limits on the DM-electron scattering cross section.
\section{Conclusion}\label{sec4}
The momentum transfer plays a crucial role in calculating the cross-section for DM scattering off electrons. If the DM has a size,  the momentum transfer can be dominated by the DM radius effect.  In this work, assuming the shape of DM particle to take the dipole form, we studied the direct detection of different radius DM via electron recoil. We found that for a large radius of DM particle, the excluded limits of the DM-electron scattering cross section is much weaker than for the point-like DM particle. For example, when the radius of DM is $\rm R_{DM}= 100~ MeV^{-1}$ and the mass of DM is sub-$\rm GeV$, the   limit on the puffy DM-electron scattering cross section is about $10^{-34}$cm$^2$ (XENON10),  $10^{-29}$cm$^2$ (XENON100)  and $ 10^{-24}$cm$^2$ (XENON1T),  which is weaker than $10^{-35}$cm$^2$ (XENON10),  $10^{-33}$cm$^2$ (XENON100)  and $ 10^{-33}$cm$^2$ (XENON1T) for the point-like DM particle. Instead, when DM particle has a very small radius, the constraints of direct detection on the DM-electron scattering cross section remain similar to the point-like DM particle case. 
 
\section*{Acknowledgements}
This work was supported by the National
Natural Science Foundation of China (NNSFC) under grant Nos. 11775012, 11821505 and 12075300,
by Peng-Huan-Wu Theoretical Physics Innovation Center (12047503), and by the Key Research Program of the Chinese Academy of Sciences, grant No. XDPB15.
%\appendix
%\renewcommand{\appendixname}{Appendix~\Alph{section}}
%\section{The  annihilation cross section in the non-relativistic situation }

\bibliographystyle{apsrev}
\bibliography{note}

\begin{thebibliography}{55}
\expandafter\ifx\csname natexlab\endcsname\relax\def\natexlab#1{#1}\fi
\expandafter\ifx\csname bibnamefont\endcsname\relax
  \def\bibnamefont#1{#1}\fi
\expandafter\ifx\csname bibfnamefont\endcsname\relax
  \def\bibfnamefont#1{#1}\fi
\expandafter\ifx\csname citenamefont\endcsname\relax
  \def\citenamefont#1{#1}\fi
\expandafter\ifx\csname url\endcsname\relax
  \def\url#1{\texttt{#1}}\fi
\expandafter\ifx\csname urlprefix\endcsname\relax\def\urlprefix{URL }\fi
\providecommand{\bibinfo}[2]{#2}
\providecommand{\eprint}[2][]{\url{#2}}

\bibitem[{\citenamefont{Aghanim et~al.}(2020)}]{Planck:2018vyg}
\bibinfo{author}{\bibfnamefont{N.}~\bibnamefont{Aghanim}} \bibnamefont{et~al.}
  (\bibinfo{collaboration}{Planck}), \bibinfo{journal}{Astron. Astrophys.}
  \textbf{\bibinfo{volume}{641}}, \bibinfo{pages}{A6} (\bibinfo{year}{2020}),
  \bibinfo{note}{[Erratum: Astron.Astrophys. 652, C4 (2021)]},
  \eprint{1807.06209}.

\bibitem[{\citenamefont{Springel et~al.}(2006)\citenamefont{Springel, Frenk,
  and White}}]{Springel:2006vs}
\bibinfo{author}{\bibfnamefont{V.}~\bibnamefont{Springel}},
  \bibinfo{author}{\bibfnamefont{C.~S.} \bibnamefont{Frenk}}, \bibnamefont{and}
  \bibinfo{author}{\bibfnamefont{S.~D.~M.} \bibnamefont{White}},
  \bibinfo{journal}{Nature} \textbf{\bibinfo{volume}{440}},
  \bibinfo{pages}{1137} (\bibinfo{year}{2006}), \eprint{astro-ph/0604561}.

\bibitem[{\citenamefont{Bahcall et~al.}(1999)\citenamefont{Bahcall, Ostriker,
  Perlmutter, and Steinhardt}}]{Bahcall:1999xn}
\bibinfo{author}{\bibfnamefont{N.~A.} \bibnamefont{Bahcall}},
  \bibinfo{author}{\bibfnamefont{J.~P.} \bibnamefont{Ostriker}},
  \bibinfo{author}{\bibfnamefont{S.}~\bibnamefont{Perlmutter}},
  \bibnamefont{and} \bibinfo{author}{\bibfnamefont{P.~J.}
  \bibnamefont{Steinhardt}}, \bibinfo{journal}{Science}
  \textbf{\bibinfo{volume}{284}}, \bibinfo{pages}{1481} (\bibinfo{year}{1999}),
  \eprint{astro-ph/9906463}.

\bibitem[{\citenamefont{Begeman et~al.}(1991)\citenamefont{Begeman, Broeils,
  and Sanders}}]{Begeman:1991iy}
\bibinfo{author}{\bibfnamefont{K.~G.} \bibnamefont{Begeman}},
  \bibinfo{author}{\bibfnamefont{A.~H.} \bibnamefont{Broeils}},
  \bibnamefont{and} \bibinfo{author}{\bibfnamefont{R.~H.}
  \bibnamefont{Sanders}}, \bibinfo{journal}{Mon. Not. Roy. Astron. Soc.}
  \textbf{\bibinfo{volume}{249}}, \bibinfo{pages}{523} (\bibinfo{year}{1991}).

\bibitem[{\citenamefont{Dyson et~al.}(1920)\citenamefont{Dyson, Eddington, and
  Davidson}}]{Dyson:1920cwa}
\bibinfo{author}{\bibfnamefont{F.~W.} \bibnamefont{Dyson}},
  \bibinfo{author}{\bibfnamefont{A.~S.} \bibnamefont{Eddington}},
  \bibnamefont{and} \bibinfo{author}{\bibfnamefont{C.}~\bibnamefont{Davidson}},
  \bibinfo{journal}{Phil. Trans. Roy. Soc. Lond. A}
  \textbf{\bibinfo{volume}{220}}, \bibinfo{pages}{291} (\bibinfo{year}{1920}).

\bibitem[{\citenamefont{Clowe et~al.}(2006)\citenamefont{Clowe, Bradac,
  Gonzalez, Markevitch, Randall, Jones, and Zaritsky}}]{Clowe:2006eq}
\bibinfo{author}{\bibfnamefont{D.}~\bibnamefont{Clowe}},
  \bibinfo{author}{\bibfnamefont{M.}~\bibnamefont{Bradac}},
  \bibinfo{author}{\bibfnamefont{A.~H.} \bibnamefont{Gonzalez}},
  \bibinfo{author}{\bibfnamefont{M.}~\bibnamefont{Markevitch}},
  \bibinfo{author}{\bibfnamefont{S.~W.} \bibnamefont{Randall}},
  \bibinfo{author}{\bibfnamefont{C.}~\bibnamefont{Jones}}, \bibnamefont{and}
  \bibinfo{author}{\bibfnamefont{D.}~\bibnamefont{Zaritsky}},
  \bibinfo{journal}{Astrophys. J. Lett.} \textbf{\bibinfo{volume}{648}},
  \bibinfo{pages}{L109} (\bibinfo{year}{2006}), \eprint{astro-ph/0608407}.

\bibitem[{\citenamefont{Lee and Weinberg}(1977)}]{Lee:1977ua}
\bibinfo{author}{\bibfnamefont{B.~W.} \bibnamefont{Lee}} \bibnamefont{and}
  \bibinfo{author}{\bibfnamefont{S.}~\bibnamefont{Weinberg}},
  \bibinfo{journal}{Phys. Rev. Lett.} \textbf{\bibinfo{volume}{39}},
  \bibinfo{pages}{165} (\bibinfo{year}{1977}).

\bibitem[{\citenamefont{Jungman et~al.}(1996)\citenamefont{Jungman,
  Kamionkowski, and Griest}}]{Jungman:1995df}
\bibinfo{author}{\bibfnamefont{G.}~\bibnamefont{Jungman}},
  \bibinfo{author}{\bibfnamefont{M.}~\bibnamefont{Kamionkowski}},
  \bibnamefont{and} \bibinfo{author}{\bibfnamefont{K.}~\bibnamefont{Griest}},
  \bibinfo{journal}{Phys. Rept.} \textbf{\bibinfo{volume}{267}},
  \bibinfo{pages}{195} (\bibinfo{year}{1996}), \eprint{hep-ph/9506380}.

\bibitem[{\citenamefont{Bertone and Tait}(2018)}]{Bertone:2018krk}
\bibinfo{author}{\bibfnamefont{G.}~\bibnamefont{Bertone}} \bibnamefont{and}
  \bibinfo{author}{\bibfnamefont{T.}~\bibnamefont{Tait}, \bibfnamefont{M.~P.}},
  \bibinfo{journal}{Nature} \textbf{\bibinfo{volume}{562}}, \bibinfo{pages}{51}
  (\bibinfo{year}{2018}), \eprint{1810.01668}.

\bibitem[{\citenamefont{Athron et~al.}(2017)}]{GAMBIT:2017zdo}
\bibinfo{author}{\bibfnamefont{P.}~\bibnamefont{Athron}} \bibnamefont{et~al.}
  (\bibinfo{collaboration}{GAMBIT}), \bibinfo{journal}{Eur. Phys. J. C}
  \textbf{\bibinfo{volume}{77}}, \bibinfo{pages}{879} (\bibinfo{year}{2017}),
  \eprint{1705.07917}.

\bibitem[{\citenamefont{Ren et~al.}(2018)}]{PandaX-II:2018xpz}
\bibinfo{author}{\bibfnamefont{X.}~\bibnamefont{Ren}} \bibnamefont{et~al.}
  (\bibinfo{collaboration}{PandaX-II}), \bibinfo{journal}{Phys. Rev. Lett.}
  \textbf{\bibinfo{volume}{121}}, \bibinfo{pages}{021304}
  (\bibinfo{year}{2018}), \eprint{1802.06912}.

\bibitem[{\citenamefont{Aprile et~al.}(2018)}]{XENON:2018voc}
\bibinfo{author}{\bibfnamefont{E.}~\bibnamefont{Aprile}} \bibnamefont{et~al.}
  (\bibinfo{collaboration}{XENON}), \bibinfo{journal}{Phys. Rev. Lett.}
  \textbf{\bibinfo{volume}{121}}, \bibinfo{pages}{111302}
  (\bibinfo{year}{2018}), \eprint{1805.12562}.

\bibitem[{\citenamefont{Akerib et~al.}(2017)}]{LUX:2017ree}
\bibinfo{author}{\bibfnamefont{D.~S.} \bibnamefont{Akerib}}
  \bibnamefont{et~al.} (\bibinfo{collaboration}{LUX}), \bibinfo{journal}{Phys.
  Rev. Lett.} \textbf{\bibinfo{volume}{118}}, \bibinfo{pages}{251302}
  (\bibinfo{year}{2017}), \eprint{1705.03380}.

\bibitem[{\citenamefont{Essig et~al.}(2016)\citenamefont{Essig,
  Fernandez-Serra, Mardon, Soto, Volansky, and Yu}}]{Essig:2015cda}
\bibinfo{author}{\bibfnamefont{R.}~\bibnamefont{Essig}},
  \bibinfo{author}{\bibfnamefont{M.}~\bibnamefont{Fernandez-Serra}},
  \bibinfo{author}{\bibfnamefont{J.}~\bibnamefont{Mardon}},
  \bibinfo{author}{\bibfnamefont{A.}~\bibnamefont{Soto}},
  \bibinfo{author}{\bibfnamefont{T.}~\bibnamefont{Volansky}}, \bibnamefont{and}
  \bibinfo{author}{\bibfnamefont{T.-T.} \bibnamefont{Yu}},
  \bibinfo{journal}{JHEP} \textbf{\bibinfo{volume}{05}}, \bibinfo{pages}{046}
  (\bibinfo{year}{2016}), \eprint{1509.01598}.

\bibitem[{\citenamefont{Essig et~al.}(2012{\natexlab{a}})\citenamefont{Essig,
  Manalaysay, Mardon, Sorensen, and Volansky}}]{Essig:2012yx}
\bibinfo{author}{\bibfnamefont{R.}~\bibnamefont{Essig}},
  \bibinfo{author}{\bibfnamefont{A.}~\bibnamefont{Manalaysay}},
  \bibinfo{author}{\bibfnamefont{J.}~\bibnamefont{Mardon}},
  \bibinfo{author}{\bibfnamefont{P.}~\bibnamefont{Sorensen}}, \bibnamefont{and}
  \bibinfo{author}{\bibfnamefont{T.}~\bibnamefont{Volansky}},
  \bibinfo{journal}{Phys. Rev. Lett.} \textbf{\bibinfo{volume}{109}},
  \bibinfo{pages}{021301} (\bibinfo{year}{2012}{\natexlab{a}}),
  \eprint{1206.2644}.

\bibitem[{\citenamefont{Essig et~al.}(2012{\natexlab{b}})\citenamefont{Essig,
  Mardon, and Volansky}}]{Essig:2011nj}
\bibinfo{author}{\bibfnamefont{R.}~\bibnamefont{Essig}},
  \bibinfo{author}{\bibfnamefont{J.}~\bibnamefont{Mardon}}, \bibnamefont{and}
  \bibinfo{author}{\bibfnamefont{T.}~\bibnamefont{Volansky}},
  \bibinfo{journal}{Phys. Rev. D} \textbf{\bibinfo{volume}{85}},
  \bibinfo{pages}{076007} (\bibinfo{year}{2012}{\natexlab{b}}),
  \eprint{1108.5383}.

\bibitem[{\citenamefont{Essig et~al.}(2017)\citenamefont{Essig, Volansky, and
  Yu}}]{Essig:2017kqs}
\bibinfo{author}{\bibfnamefont{R.}~\bibnamefont{Essig}},
  \bibinfo{author}{\bibfnamefont{T.}~\bibnamefont{Volansky}}, \bibnamefont{and}
  \bibinfo{author}{\bibfnamefont{T.-T.} \bibnamefont{Yu}},
  \bibinfo{journal}{Phys. Rev. D} \textbf{\bibinfo{volume}{96}},
  \bibinfo{pages}{043017} (\bibinfo{year}{2017}), \eprint{1703.00910}.

\bibitem[{\citenamefont{Chen et~al.}(2015)\citenamefont{Chen, Chi, Liu, Wu, and
  Wu}}]{Chen:2015pha}
\bibinfo{author}{\bibfnamefont{J.-W.} \bibnamefont{Chen}},
  \bibinfo{author}{\bibfnamefont{H.-C.} \bibnamefont{Chi}},
  \bibinfo{author}{\bibfnamefont{C.~P.} \bibnamefont{Liu}},
  \bibinfo{author}{\bibfnamefont{C.-L.} \bibnamefont{Wu}}, \bibnamefont{and}
  \bibinfo{author}{\bibfnamefont{C.-P.} \bibnamefont{Wu}},
  \bibinfo{journal}{Phys. Rev. D} \textbf{\bibinfo{volume}{92}},
  \bibinfo{pages}{096013} (\bibinfo{year}{2015}), \eprint{1508.03508}.

\bibitem[{\citenamefont{Bloch et~al.}(2021)\citenamefont{Bloch, Caputo, Essig,
  Redigolo, Sholapurkar, and Volansky}}]{Bloch:2020uzh}
\bibinfo{author}{\bibfnamefont{I.~M.} \bibnamefont{Bloch}},
  \bibinfo{author}{\bibfnamefont{A.}~\bibnamefont{Caputo}},
  \bibinfo{author}{\bibfnamefont{R.}~\bibnamefont{Essig}},
  \bibinfo{author}{\bibfnamefont{D.}~\bibnamefont{Redigolo}},
  \bibinfo{author}{\bibfnamefont{M.}~\bibnamefont{Sholapurkar}},
  \bibnamefont{and} \bibinfo{author}{\bibfnamefont{T.}~\bibnamefont{Volansky}},
  \bibinfo{journal}{JHEP} \textbf{\bibinfo{volume}{01}}, \bibinfo{pages}{178}
  (\bibinfo{year}{2021}), \eprint{2006.14521}.

\bibitem[{\citenamefont{Derenzo et~al.}(2017)\citenamefont{Derenzo, Essig,
  Massari, Soto, and Yu}}]{Derenzo:2016fse}
\bibinfo{author}{\bibfnamefont{S.}~\bibnamefont{Derenzo}},
  \bibinfo{author}{\bibfnamefont{R.}~\bibnamefont{Essig}},
  \bibinfo{author}{\bibfnamefont{A.}~\bibnamefont{Massari}},
  \bibinfo{author}{\bibfnamefont{A.}~\bibnamefont{Soto}}, \bibnamefont{and}
  \bibinfo{author}{\bibfnamefont{T.-T.} \bibnamefont{Yu}},
  \bibinfo{journal}{Phys. Rev. D} \textbf{\bibinfo{volume}{96}},
  \bibinfo{pages}{016026} (\bibinfo{year}{2017}), \eprint{1607.01009}.

\bibitem[{\citenamefont{Blanco et~al.}(2020)\citenamefont{Blanco, Collar, Kahn,
  and Lillard}}]{Blanco:2019lrf}
\bibinfo{author}{\bibfnamefont{C.}~\bibnamefont{Blanco}},
  \bibinfo{author}{\bibfnamefont{J.~I.} \bibnamefont{Collar}},
  \bibinfo{author}{\bibfnamefont{Y.}~\bibnamefont{Kahn}}, \bibnamefont{and}
  \bibinfo{author}{\bibfnamefont{B.}~\bibnamefont{Lillard}},
  \bibinfo{journal}{Phys. Rev. D} \textbf{\bibinfo{volume}{101}},
  \bibinfo{pages}{056001} (\bibinfo{year}{2020}), \eprint{1912.02822}.

\bibitem[{\citenamefont{Pandey et~al.}(2020)\citenamefont{Pandey, Singh, Wu,
  Chen, Chi, Hsieh, Liu, and Wong}}]{Pandey:2018esq}
\bibinfo{author}{\bibfnamefont{M.~K.} \bibnamefont{Pandey}},
  \bibinfo{author}{\bibfnamefont{L.}~\bibnamefont{Singh}},
  \bibinfo{author}{\bibfnamefont{C.-P.} \bibnamefont{Wu}},
  \bibinfo{author}{\bibfnamefont{J.-W.} \bibnamefont{Chen}},
  \bibinfo{author}{\bibfnamefont{H.-C.} \bibnamefont{Chi}},
  \bibinfo{author}{\bibfnamefont{C.-C.} \bibnamefont{Hsieh}},
  \bibinfo{author}{\bibfnamefont{C.~P.} \bibnamefont{Liu}}, \bibnamefont{and}
  \bibinfo{author}{\bibfnamefont{H.~T.} \bibnamefont{Wong}},
  \bibinfo{journal}{Phys. Rev. D} \textbf{\bibinfo{volume}{102}},
  \bibinfo{pages}{123025} (\bibinfo{year}{2020}), \eprint{1812.11759}.

\bibitem[{\citenamefont{Liu et~al.}(2022)\citenamefont{Liu, Wu, Chen, Chi,
  Pandey, Singh, and Wong}}]{Liu:2021avx}
\bibinfo{author}{\bibfnamefont{C.~P.} \bibnamefont{Liu}},
  \bibinfo{author}{\bibfnamefont{C.-P.} \bibnamefont{Wu}},
  \bibinfo{author}{\bibfnamefont{J.-W.} \bibnamefont{Chen}},
  \bibinfo{author}{\bibfnamefont{H.-C.} \bibnamefont{Chi}},
  \bibinfo{author}{\bibfnamefont{M.~K.} \bibnamefont{Pandey}},
  \bibinfo{author}{\bibfnamefont{L.}~\bibnamefont{Singh}}, \bibnamefont{and}
  \bibinfo{author}{\bibfnamefont{H.~T.} \bibnamefont{Wong}},
  \bibinfo{journal}{Phys. Rev. D} \textbf{\bibinfo{volume}{106}},
  \bibinfo{pages}{063003} (\bibinfo{year}{2022}), \eprint{2106.16214}.

\bibitem[{\citenamefont{Aprile et~al.}(2022)}]{XENONCollaborationSS:2021sgk}
\bibinfo{author}{\bibfnamefont{E.}~\bibnamefont{Aprile}} \bibnamefont{et~al.}
  (\bibinfo{collaboration}{(XENON Collaboration)\textsection{}, XENON}),
  \bibinfo{journal}{Phys. Rev. D} \textbf{\bibinfo{volume}{106}},
  \bibinfo{pages}{022001} (\bibinfo{year}{2022}), \eprint{2112.12116}.

\bibitem[{\citenamefont{Chavarria}(2022)}]{Chavarria:2022par}
\bibinfo{author}{\bibfnamefont{A.~E.} \bibnamefont{Chavarria}}, in
  \emph{\bibinfo{booktitle}{{14th International Workshop on the Identification
  of Dark Matter 2022}}} (\bibinfo{year}{2022}), \eprint{2210.05661}.

\bibitem[{\citenamefont{Ibe et~al.}(2018)\citenamefont{Ibe, Nakano, Shoji, and
  Suzuki}}]{Ibe:2017yqa}
\bibinfo{author}{\bibfnamefont{M.}~\bibnamefont{Ibe}},
  \bibinfo{author}{\bibfnamefont{W.}~\bibnamefont{Nakano}},
  \bibinfo{author}{\bibfnamefont{Y.}~\bibnamefont{Shoji}}, \bibnamefont{and}
  \bibinfo{author}{\bibfnamefont{K.}~\bibnamefont{Suzuki}},
  \bibinfo{journal}{JHEP} \textbf{\bibinfo{volume}{03}}, \bibinfo{pages}{194}
  (\bibinfo{year}{2018}), \eprint{1707.07258}.

\bibitem[{\citenamefont{Essig et~al.}(2020)\citenamefont{Essig, Pradler,
  Sholapurkar, and Yu}}]{Essig:2019xkx}
\bibinfo{author}{\bibfnamefont{R.}~\bibnamefont{Essig}},
  \bibinfo{author}{\bibfnamefont{J.}~\bibnamefont{Pradler}},
  \bibinfo{author}{\bibfnamefont{M.}~\bibnamefont{Sholapurkar}},
  \bibnamefont{and} \bibinfo{author}{\bibfnamefont{T.-T.} \bibnamefont{Yu}},
  \bibinfo{journal}{Phys. Rev. Lett.} \textbf{\bibinfo{volume}{124}},
  \bibinfo{pages}{021801} (\bibinfo{year}{2020}), \eprint{1908.10881}.

\bibitem[{\citenamefont{Wang et~al.}(2022{\natexlab{a}})\citenamefont{Wang, Wu,
  Wu, and Zhu}}]{Wang:2021oha}
\bibinfo{author}{\bibfnamefont{W.}~\bibnamefont{Wang}},
  \bibinfo{author}{\bibfnamefont{K.-Y.} \bibnamefont{Wu}},
  \bibinfo{author}{\bibfnamefont{L.}~\bibnamefont{Wu}}, \bibnamefont{and}
  \bibinfo{author}{\bibfnamefont{B.}~\bibnamefont{Zhu}},
  \bibinfo{journal}{Nucl. Phys. B} \textbf{\bibinfo{volume}{983}},
  \bibinfo{pages}{115907} (\bibinfo{year}{2022}{\natexlab{a}}),
  \eprint{2112.06492}.

\bibitem[{\citenamefont{Knapen et~al.}(2021)\citenamefont{Knapen, Kozaczuk, and
  Lin}}]{Knapen:2020aky}
\bibinfo{author}{\bibfnamefont{S.}~\bibnamefont{Knapen}},
  \bibinfo{author}{\bibfnamefont{J.}~\bibnamefont{Kozaczuk}}, \bibnamefont{and}
  \bibinfo{author}{\bibfnamefont{T.}~\bibnamefont{Lin}},
  \bibinfo{journal}{Phys. Rev. Lett.} \textbf{\bibinfo{volume}{127}},
  \bibinfo{pages}{081805} (\bibinfo{year}{2021}), \eprint{2011.09496}.

\bibitem[{\citenamefont{Liang et~al.}(2021)\citenamefont{Liang, Mo, Zheng, and
  Zhang}}]{Liang:2020ryg}
\bibinfo{author}{\bibfnamefont{Z.-L.} \bibnamefont{Liang}},
  \bibinfo{author}{\bibfnamefont{C.}~\bibnamefont{Mo}},
  \bibinfo{author}{\bibfnamefont{F.}~\bibnamefont{Zheng}}, \bibnamefont{and}
  \bibinfo{author}{\bibfnamefont{P.}~\bibnamefont{Zhang}},
  \bibinfo{journal}{Phys. Rev. D} \textbf{\bibinfo{volume}{104}},
  \bibinfo{pages}{056009} (\bibinfo{year}{2021}), \eprint{2011.13352}.

\bibitem[{\citenamefont{Chu et~al.}(2020{\natexlab{a}})\citenamefont{Chu,
  Garcia-Cely, and Murayama}}]{Chu:2018faw}
\bibinfo{author}{\bibfnamefont{X.}~\bibnamefont{Chu}},
  \bibinfo{author}{\bibfnamefont{C.}~\bibnamefont{Garcia-Cely}},
  \bibnamefont{and} \bibinfo{author}{\bibfnamefont{H.}~\bibnamefont{Murayama}},
  \bibinfo{journal}{Phys. Rev. Lett.} \textbf{\bibinfo{volume}{124}},
  \bibinfo{pages}{041101} (\bibinfo{year}{2020}{\natexlab{a}}),
  \eprint{1901.00075}.

\bibitem[{\citenamefont{Wang et~al.}(2023{\natexlab{a}})\citenamefont{Wang, Xu,
  Yang, and Zhu}}]{Wang:2023xii}
\bibinfo{author}{\bibfnamefont{W.}~\bibnamefont{Wang}},
  \bibinfo{author}{\bibfnamefont{W.-L.} \bibnamefont{Xu}},
  \bibinfo{author}{\bibfnamefont{J.~M.} \bibnamefont{Yang}}, \bibnamefont{and}
  \bibinfo{author}{\bibfnamefont{B.}~\bibnamefont{Zhu}}
  (\bibinfo{year}{2023}{\natexlab{a}}), \eprint{2303.11058}.

\bibitem[{\citenamefont{Wang et~al.}(2022{\natexlab{b}})\citenamefont{Wang, Xu,
  and Zhu}}]{Wang:2021tjf}
\bibinfo{author}{\bibfnamefont{W.}~\bibnamefont{Wang}},
  \bibinfo{author}{\bibfnamefont{W.-L.} \bibnamefont{Xu}}, \bibnamefont{and}
  \bibinfo{author}{\bibfnamefont{B.}~\bibnamefont{Zhu}},
  \bibinfo{journal}{Phys. Rev. D} \textbf{\bibinfo{volume}{105}},
  \bibinfo{pages}{075013} (\bibinfo{year}{2022}{\natexlab{b}}),
  \eprint{2108.07030}.

\bibitem[{\citenamefont{Chu et~al.}(2020{\natexlab{b}})\citenamefont{Chu,
  Garcia-Cely, and Murayama}}]{Chu:2019awd}
\bibinfo{author}{\bibfnamefont{X.}~\bibnamefont{Chu}},
  \bibinfo{author}{\bibfnamefont{C.}~\bibnamefont{Garcia-Cely}},
  \bibnamefont{and} \bibinfo{author}{\bibfnamefont{H.}~\bibnamefont{Murayama}},
  \bibinfo{journal}{JCAP} \textbf{\bibinfo{volume}{06}}, \bibinfo{pages}{043}
  (\bibinfo{year}{2020}{\natexlab{b}}), \eprint{1908.06067}.

\bibitem[{\citenamefont{Laha}(2015)}]{Laha:2015yoa}
\bibinfo{author}{\bibfnamefont{R.}~\bibnamefont{Laha}}, \bibinfo{journal}{Phys.
  Rev. D} \textbf{\bibinfo{volume}{92}}, \bibinfo{pages}{083509}
  (\bibinfo{year}{2015}), \eprint{1505.02772}.

\bibitem[{\citenamefont{Laha and Braaten}(2014)}]{Laha:2013gva}
\bibinfo{author}{\bibfnamefont{R.}~\bibnamefont{Laha}} \bibnamefont{and}
  \bibinfo{author}{\bibfnamefont{E.}~\bibnamefont{Braaten}},
  \bibinfo{journal}{Phys. Rev. D} \textbf{\bibinfo{volume}{89}},
  \bibinfo{pages}{103510} (\bibinfo{year}{2014}), \eprint{1311.6386}.

\bibitem[{\citenamefont{Cline}(2022)}]{Cline:2021itd}
\bibinfo{author}{\bibfnamefont{J.~M.} \bibnamefont{Cline}},
  \bibinfo{journal}{SciPost Phys. Lect. Notes} \textbf{\bibinfo{volume}{52}},
  \bibinfo{pages}{1} (\bibinfo{year}{2022}), \eprint{2108.10314}.

\bibitem[{\citenamefont{Tsai et~al.}(2022)\citenamefont{Tsai, McGehee, and
  Murayama}}]{Tsai:2020vpi}
\bibinfo{author}{\bibfnamefont{Y.-D.} \bibnamefont{Tsai}},
  \bibinfo{author}{\bibfnamefont{R.}~\bibnamefont{McGehee}}, \bibnamefont{and}
  \bibinfo{author}{\bibfnamefont{H.}~\bibnamefont{Murayama}},
  \bibinfo{journal}{Phys. Rev. Lett.} \textbf{\bibinfo{volume}{128}},
  \bibinfo{pages}{172001} (\bibinfo{year}{2022}), \eprint{2008.08608}.

\bibitem[{\citenamefont{Hardy et~al.}(2015)\citenamefont{Hardy, Lasenby,
  March-Russell, and West}}]{Hardy:2015boa}
\bibinfo{author}{\bibfnamefont{E.}~\bibnamefont{Hardy}},
  \bibinfo{author}{\bibfnamefont{R.}~\bibnamefont{Lasenby}},
  \bibinfo{author}{\bibfnamefont{J.}~\bibnamefont{March-Russell}},
  \bibnamefont{and} \bibinfo{author}{\bibfnamefont{S.~M.} \bibnamefont{West}},
  \bibinfo{journal}{JHEP} \textbf{\bibinfo{volume}{07}}, \bibinfo{pages}{133}
  (\bibinfo{year}{2015}), \eprint{1504.05419}.

\bibitem[{\citenamefont{Acevedo et~al.}(2022)\citenamefont{Acevedo, Bramante,
  and Goodman}}]{Acevedo:2021kly}
\bibinfo{author}{\bibfnamefont{J.~F.} \bibnamefont{Acevedo}},
  \bibinfo{author}{\bibfnamefont{J.}~\bibnamefont{Bramante}}, \bibnamefont{and}
  \bibinfo{author}{\bibfnamefont{A.}~\bibnamefont{Goodman}},
  \bibinfo{journal}{Phys. Rev. D} \textbf{\bibinfo{volume}{105}},
  \bibinfo{pages}{023012} (\bibinfo{year}{2022}), \eprint{2108.10889}.

\bibitem[{\citenamefont{Engel}(1991)}]{Engel:1991wq}
\bibinfo{author}{\bibfnamefont{J.}~\bibnamefont{Engel}},
  \bibinfo{journal}{Phys. Lett. B} \textbf{\bibinfo{volume}{264}},
  \bibinfo{pages}{114} (\bibinfo{year}{1991}).

\bibitem[{\citenamefont{Feldstein et~al.}(2010)\citenamefont{Feldstein,
  Fitzpatrick, and Katz}}]{Feldstein:2009tr}
\bibinfo{author}{\bibfnamefont{B.}~\bibnamefont{Feldstein}},
  \bibinfo{author}{\bibfnamefont{A.~L.} \bibnamefont{Fitzpatrick}},
  \bibnamefont{and} \bibinfo{author}{\bibfnamefont{E.}~\bibnamefont{Katz}},
  \bibinfo{journal}{JCAP} \textbf{\bibinfo{volume}{01}}, \bibinfo{pages}{020}
  (\bibinfo{year}{2010}), \eprint{0908.2991}.

\bibitem[{\citenamefont{Wang et~al.}(2023{\natexlab{b}})\citenamefont{Wang, Xu,
  Yang, and Zhu}}]{Wang:2023wrx}
\bibinfo{author}{\bibfnamefont{W.}~\bibnamefont{Wang}},
  \bibinfo{author}{\bibfnamefont{W.-L.} \bibnamefont{Xu}},
  \bibinfo{author}{\bibfnamefont{J.~M.} \bibnamefont{Yang}}, \bibnamefont{and}
  \bibinfo{author}{\bibfnamefont{R.}~\bibnamefont{Zhu}}
  (\bibinfo{year}{2023}{\natexlab{b}}), \eprint{2305.12668}.

\bibitem[{\citenamefont{Bunge et~al.}(1993)\citenamefont{Bunge, Barrientos, and
  Bunge}}]{Bunge:1993jsz}
\bibinfo{author}{\bibfnamefont{C.~F.} \bibnamefont{Bunge}},
  \bibinfo{author}{\bibfnamefont{J.~A.} \bibnamefont{Barrientos}},
  \bibnamefont{and} \bibinfo{author}{\bibfnamefont{A.~V.} \bibnamefont{Bunge}},
  \bibinfo{journal}{Atom. Data Nucl. Data Tabl.} \textbf{\bibinfo{volume}{53}},
  \bibinfo{pages}{113} (\bibinfo{year}{1993}).

\bibitem[{\citenamefont{Catena et~al.}(2020)\citenamefont{Catena, Emken,
  Spaldin, and Tarantino}}]{Catena:2019gfa}
\bibinfo{author}{\bibfnamefont{R.}~\bibnamefont{Catena}},
  \bibinfo{author}{\bibfnamefont{T.}~\bibnamefont{Emken}},
  \bibinfo{author}{\bibfnamefont{N.~A.} \bibnamefont{Spaldin}},
  \bibnamefont{and}
  \bibinfo{author}{\bibfnamefont{W.}~\bibnamefont{Tarantino}},
  \bibinfo{journal}{Phys. Rev. Res.} \textbf{\bibinfo{volume}{2}},
  \bibinfo{pages}{033195} (\bibinfo{year}{2020}), \eprint{1912.08204}.

\bibitem[{\citenamefont{Griffin et~al.}(2021)\citenamefont{Griffin, Inzani,
  Trickle, Zhang, and Zurek}}]{Griffin:2021znd}
\bibinfo{author}{\bibfnamefont{S.~M.} \bibnamefont{Griffin}},
  \bibinfo{author}{\bibfnamefont{K.}~\bibnamefont{Inzani}},
  \bibinfo{author}{\bibfnamefont{T.}~\bibnamefont{Trickle}},
  \bibinfo{author}{\bibfnamefont{Z.}~\bibnamefont{Zhang}}, \bibnamefont{and}
  \bibinfo{author}{\bibfnamefont{K.~M.} \bibnamefont{Zurek}},
  \bibinfo{journal}{Phys. Rev. D} \textbf{\bibinfo{volume}{104}},
  \bibinfo{pages}{095015} (\bibinfo{year}{2021}), \eprint{2105.05253}.

\bibitem[{\citenamefont{Maity et~al.}(2021)\citenamefont{Maity, Ray, and
  Sarkar}}]{Maity:2020wic}
\bibinfo{author}{\bibfnamefont{T.~N.} \bibnamefont{Maity}},
  \bibinfo{author}{\bibfnamefont{T.~S.} \bibnamefont{Ray}}, \bibnamefont{and}
  \bibinfo{author}{\bibfnamefont{S.}~\bibnamefont{Sarkar}},
  \bibinfo{journal}{Eur. Phys. J. C} \textbf{\bibinfo{volume}{81}},
  \bibinfo{pages}{1005} (\bibinfo{year}{2021}), \eprint{2011.12896}.

\bibitem[{\citenamefont{Radick et~al.}(2021)\citenamefont{Radick, Taki, and
  Yu}}]{Radick:2020qip}
\bibinfo{author}{\bibfnamefont{A.}~\bibnamefont{Radick}},
  \bibinfo{author}{\bibfnamefont{A.-M.} \bibnamefont{Taki}}, \bibnamefont{and}
  \bibinfo{author}{\bibfnamefont{T.-T.} \bibnamefont{Yu}},
  \bibinfo{journal}{JCAP} \textbf{\bibinfo{volume}{02}}, \bibinfo{pages}{004}
  (\bibinfo{year}{2021}), \eprint{2011.02493}.

\bibitem[{\citenamefont{Savage et~al.}(2009)\citenamefont{Savage, Gelmini,
  Gondolo, and Freese}}]{Savage:2008er}
\bibinfo{author}{\bibfnamefont{C.}~\bibnamefont{Savage}},
  \bibinfo{author}{\bibfnamefont{G.}~\bibnamefont{Gelmini}},
  \bibinfo{author}{\bibfnamefont{P.}~\bibnamefont{Gondolo}}, \bibnamefont{and}
  \bibinfo{author}{\bibfnamefont{K.}~\bibnamefont{Freese}},
  \bibinfo{journal}{JCAP} \textbf{\bibinfo{volume}{04}}, \bibinfo{pages}{010}
  (\bibinfo{year}{2009}), \eprint{0808.3607}.

\bibitem[{\citenamefont{Dehnen and Binney}(1998)}]{Dehnen:1997cq}
\bibinfo{author}{\bibfnamefont{W.}~\bibnamefont{Dehnen}} \bibnamefont{and}
  \bibinfo{author}{\bibfnamefont{J.}~\bibnamefont{Binney}},
  \bibinfo{journal}{Mon. Not. Roy. Astron. Soc.}
  \textbf{\bibinfo{volume}{298}}, \bibinfo{pages}{387} (\bibinfo{year}{1998}),
  \eprint{astro-ph/9710077}.

\bibitem[{\citenamefont{Smith et~al.}(2007)}]{Smith:2006ym}
\bibinfo{author}{\bibfnamefont{M.~C.} \bibnamefont{Smith}}
  \bibnamefont{et~al.}, \bibinfo{journal}{Mon. Not. Roy. Astron. Soc.}
  \textbf{\bibinfo{volume}{379}}, \bibinfo{pages}{755} (\bibinfo{year}{2007}),
  \eprint{astro-ph/0611671}.

\bibitem[{\citenamefont{Shutt et~al.}(2007)\citenamefont{Shutt, Dahl, Kwong,
  Bolozdynya, and Brusov}}]{Shutt:2006ed}
\bibinfo{author}{\bibfnamefont{T.}~\bibnamefont{Shutt}},
  \bibinfo{author}{\bibfnamefont{C.~E.} \bibnamefont{Dahl}},
  \bibinfo{author}{\bibfnamefont{J.}~\bibnamefont{Kwong}},
  \bibinfo{author}{\bibfnamefont{A.}~\bibnamefont{Bolozdynya}},
  \bibnamefont{and} \bibinfo{author}{\bibfnamefont{P.}~\bibnamefont{Brusov}},
  \bibinfo{journal}{Nucl. Instrum. Meth. A} \textbf{\bibinfo{volume}{579}},
  \bibinfo{pages}{451} (\bibinfo{year}{2007}), \eprint{astro-ph/0608137}.

\bibitem[{\citenamefont{Aprile et~al.}(2012)}]{XENON100:2011cza}
\bibinfo{author}{\bibfnamefont{E.}~\bibnamefont{Aprile}} \bibnamefont{et~al.}
  (\bibinfo{collaboration}{XENON100}), \bibinfo{journal}{Astropart. Phys.}
  \textbf{\bibinfo{volume}{35}}, \bibinfo{pages}{573} (\bibinfo{year}{2012}),
  \eprint{1107.2155}.

\bibitem[{\citenamefont{Aprile et~al.}(2019)}]{XENON:2019gfn}
\bibinfo{author}{\bibfnamefont{E.}~\bibnamefont{Aprile}} \bibnamefont{et~al.}
  (\bibinfo{collaboration}{XENON}), \bibinfo{journal}{Phys. Rev. Lett.}
  \textbf{\bibinfo{volume}{123}}, \bibinfo{pages}{251801}
  (\bibinfo{year}{2019}), \eprint{1907.11485}.

\bibitem[{\citenamefont{Aprile et~al.}(2016)}]{XENON:2016jmt}
\bibinfo{author}{\bibfnamefont{E.}~\bibnamefont{Aprile}} \bibnamefont{et~al.}
  (\bibinfo{collaboration}{XENON}), \bibinfo{journal}{Phys. Rev. D}
  \textbf{\bibinfo{volume}{94}}, \bibinfo{pages}{092001}
  (\bibinfo{year}{2016}), \bibinfo{note}{[Erratum: Phys.Rev.D 95, 059901
  (2017)]}, \eprint{1605.06262}.

\end{thebibliography}

\end{document}